\documentclass[pra]{revtex4}
\usepackage{amssymb}
\usepackage{amsmath}
\usepackage{graphics}
\newcommand{\vc}[1]{\boldsymbol{#1}}

\begin{document}

\title{ Hamiltonian for a particle in a magnetic field on a curved surface in orthogonal curvilinear coordinates }

\author{M. S. Shikakhwa}
\affiliation{Department of physics, The University of Jordan,Amman 11942 Jordan
and Middle East Technical University Northern Cyprus Campus,Kalkanl\i, G\"{u}zelyurt, via Mersin 10, Turkey}
\author{N. Chair}
\affiliation{Department of physics, The University of Jordan,Amman 11942 Jordan}

\begin{abstract}
The  Schr\"{o}dinger Hamiltonian of a spin zero particle as well as the Pauli Hamiltonian with spin-orbit coupling included of a spin one-half particle in  electromagnetic fields that are confined to a curved surface  embedded in a three-dimensional space spanned by  a general Orthogonal Curvilinear Coordinate (OCC) are constructed. A new approach, based on the physical argument that upon squeezing  the particle to the surface by a potential, then it is the physical gauge-covariant kinematical momentum operator (velocity operator) transverse to the surface that should be dropped from the Hamiltonian(s). In both cases,the resulting Hermitian gauge-invariant Hamiltonian on the surface is free from any reference to the component of the vector potential transverse to the surface, and the approach is completely gauge-independent. In particular, for the Pauli Hamiltonian these results are obtained exactly without any further assumptions or approximations. Explicit covariant plug-and-play formulae for the Schr\"{o}dinger Hamiltonians on the surfaces of a cylinder, a sphere and a torus are derived.
\end{abstract}

\maketitle

\section{Introduction}
The recent technological advances leading to the synthesis of nano-scale curved structures, like nanotubes and nanobubbles have boosted the interest in the quantum mechanics of particles confined to 2D curved surfaces. A major approach \cite{Koppe,Costa} to this problem is the so called thin-layer quantization procedure (as examples of other approaches, see the works \cite{ghosh1,ghosh2}). The idea is to embed the 2D surface into the larger full 3D Euclidean space and achieve confinement of the particle to the surface by introducing a squeezing potential. More specifically, one considers a curvilinear coordinate system with coordinates $q_1$ and $q_2$ at the surface, and the coordinate $q_3$ in its vicinity in the direction normal to it. The position vector is thus written as $\vc{r}(q_1,q_2,q_3)=\vc{r}_S(q_1,q_2)+\vc{\hat{n}}q_3$ , where  $\vc{\hat{n}}$ is a unit vector normal to the surface. The Schr\"{o}dinger equation for a spin zero particle is written in terms of these variables, and the limit $q_3\rightarrow 0$ for a sufficiently strong squeezing potential $V(q_3)$ is taken. The Hamiltonian  then reduces to the sum of two independent on-surface and transverse parts, with the latter containing only the transverse, i.e. the 3-dynamics. This transverse Hamiltonian is then dropped on the ground that the transverse excitations for a sufficiently strong confining potential have a much higher energy than those at the surface, and so can be safely neglected in comparison to the range of energies considered.This way,one achieves decoupling of the transverse  dynamics and is left  with only the surface Hamiltonian. For an otherwise free spin zero particle, this mechanism generates a geometric kinetic energy term, the optical analogue of which has recently been observed \cite{optical GKE}. \\
The natural extension of the thin-layer quantization to the case of a spin zero particle in an electromagnetic field was recently considered in a series of works \cite{Encinosa,ferrari,Jensen1,Jensen2,Ortix1} reporting various and sometime even contradictory results. The work \cite{Encinosa} reported that when thin-layer quantization was applied to a particle confined to a cylindrical surface in an arbitrary constant magnetic field, the Hamiltonian contained a term that coupled the transverse component of the vector potential to the mean curvature of the surface. In \cite{ferrari}, it was argued that there was no such coupling; it was reported,however,that the surface and transverse dynamics could be decoupled only upon employing a gauge transformation that took the  transverse component of the vector potential to zero. It was argued later in \cite{Jensen1,Jensen2} that this coupling-to-the-curvature term indeed appeared in the Hamiltonian. However, it was shown in \cite{Ortix1} that this term  dropped once the analysis were carried out carefully . The fact that one had yet to consider a gauge transformation to remove the transverse component of the vector potential  from the theory was also confirmed in this work.\\
 Applying the thin-layer quantization to the Pauli equation for a spin one-half particle in an electromagnetic field was the second natural extension of the formalism \cite{Wang,Kosugi}.In \cite{Wang} the same conclusions as in \cite{ferrari} were reached. In \cite{Kosugi} the Pauli Hamiltonian in the presence of an electromagnetic field plus relativistic corrections; the spin-orbit coupling (SOC) was  considered. It was found that decoupling of the transverse dynamics in this case can be achieved only under certain assumptions and approximations. Hamiltonians of spin one-half particles on curved surfaces with specific SOC relevant to condensed matter physics, namely the Rashba \cite{rashba} and Dresselhaus \cite{dresselhaus} SOC's in the absence of magnetic fields were also considered \cite{Entin and Magaril,u shape,cheng,exact,shikakhwa and chair}. In these works, a new geometric potential appeared upon confining the particle to the surface that was essential in this case to render the Hamiltonian(s) Hermitian.

The present work  extends a new, simple and physics-based approach recently introduced in \cite{shikakhwa and chair}, to construct the Hamiltonian of a particle in an electromagnetic field on a curved surface embedded in a 3D space spanned by an \textit{orthogonal} curvilinear coordinate (OCC) system . The Hamiltonians of both spin zero and spin one-half particles, with SOC  present in the latter, are constructed. In this approach, the 'decoupling of the transverse dynamics' from the Hamiltonian in the limit $q_3\rightarrow 0$ is interpreted on physical grounds as amounting to dropping the transverse gauge-covariant kinematical momentum from the Hamiltonian. Upon doing so,the resulting Hermitian surface Hamiltonian is completely free from any reference to the transverse component of the vector potential without the need to a further gauge transformation, and the geometric kinetic energy term emerges naturally. In the case of SOC, the well-known geometric potential also appears and the resulting SOC term is gauge-covariant,too, without the need for any assumptions or approximations. While it is true that confining the treatment to OCC limits the generality of the results, the transparency of the physics and the intuition gained, in addition to the fact that  the  formalism developed for general curvilinear coordinates is in many cases applied to surfaces that can be embedded in OCC system, e.g. spheres, cylinders and tori provides a sufficient motivation for the use of OCC. Moreover, in the last paragraph of the Conclusions, it is argued that our approach can be also applied to genreal -not necessarily- curvilinear coordinates.\\

 In section 2, the general formalism is introduced and applied to the simplest case of a spin-zero particle confined to a curved surface. Section 3 considers spin zero particle in a static electromagnetic field. The general expression for the Hamiltonian constructed is applied to the three geometries; the surfaces of a cylinder, a sphere and a torus.The case of Pauli Hamiltonian in an electromagnetic field with SOC is the subject of section 4. Discussion and conclusions are given in section 5.\\

\section{ Hamiltonian of a spin zero particle  on a curved surface  in OCC's}

The Hamiltonian for a free-particle in 3D expressed in general OCC denoted $q_1,q_2$ and $q_3$ is given by:
\begin{equation}\label{laplacian}
 H =\frac{-\hbar^2}{2m}\nabla^2=\frac{-\hbar^2}{2m}(\frac{1}{h_1 h_2 h_3})(\partial_1\frac{h_2 h_3}{h_1}\partial_1+\partial_2\frac{h_1 h_3}{h_2}\partial_2+\partial_3\frac{h_1 h_2}{h_3}\partial_3)
\end{equation}
$h_1 ,h_2$ and $h_3$ are the well-known \cite{Arfken}scale factors of the OCC's defined in terms of the derivatives of the position vector $\vc{r}$ as $\frac{\partial\vc{r}}{\partial q_i}=h_i\vc{\hat{e}}_i$, where $\vc{\hat{e}}_i$'s are the orthogonal unit vectors of the OCC system, and we use $\partial_i\equiv\frac{\partial}{\partial q_i}$. The task to be achieved is writing down the Hamiltonian for the particle when it is confined to a surface embedded in the 3D space spanned by the above OCC. Such a surface is defined in the space spanned by the OCC  by setting  one coordinate that has  the dimensions of length; $q_3$ say,( thus has a  scale factor $h_3=1$ )  to a constant,e.g. $q_3=a$. Confining the particle to this surface can then be achieved by introducing a deep confining potential $V(q_3)$ - not shown in the above Hamiltonian-  that squeezes the particle into a thin layer around $a$ and taking the limit $q_3\rightarrow a$. Now, we follow a new physics-based practical approach we recently introduced \cite{shikakhwa and chair} to generate the correct Hermitian Hamiltonian on the surface. First, we find the  Hermitian canonical momentum conjugate to $q_3$ which  is not simply $-i\hbar\partial_3$ as this latter is not Hermitian in the sense \cite{Griffiths}

 \begin{equation*}\label{}
\langle \Psi|-i\hbar\partial_3\Psi\rangle=\langle -i\hbar(\partial_3+\frac{1}{h_1 h_2 h_3}\partial_3(h_1 h_2))\Psi|\Psi\rangle\neq\langle -i\hbar\partial_3\Psi|\Psi\rangle
\end{equation*}
where integration is over all space with the  measure $d^3 q h_1 h_2 h_3$ and a surface term was dropped as usual.
We can check that the following operator is indeed Hermitian and thus can be taken as the canonical momentum conjugate to $q_3$ (recall that $h_3=1$):
\begin{equation}\label{hermitian p}
 p_3=-i\hbar(\frac{1}{h_3}\partial_3+\frac{1}{2h_1 h_2 h_3}\partial_3(h_1 h_2))
\end{equation}
 The Laplacian operator can then be rewritten as:
\begin{equation*}\label{p3 square}
 \frac{-\hbar^2}{2m}\nabla^2=\frac{p_3^2}{2m}+\frac{\hbar^2}{2m}(\frac{\partial_3^2(h_1 h_2)}{2h_1 h_2}-\frac{(\partial_3(h_1 h_2))^2}{(2h_1 h_2)^2})+\frac{-\hbar^2}{2m}(\frac{1}{h_1 h_2 h_3})(\partial_1\frac{h_2 h_3}{h_1}\partial_1+\partial_2\frac{h_1 h_3}{h_2}\partial_2)
\end{equation*}
Now, it has been proven in \cite{shikakhwa and chair} that the second term in the above expression is nothing but the famous geometric kinetic energy :
\begin{equation}\label{geometric K}
\frac{\hbar^2}{2m}(\frac{\partial_3^2(h_1 h_2)}{2h_1 h_2}-\frac{(\partial_3(h_1 h_2))^2}{(2h_1 h_2)^2})=-\frac{\hbar^2}{2m}(M^2-K)
\end{equation}
where $M$ and $K$ are, respectively, the mean and the Gaussian curvatures of the surface. Thus, the OCC Hamiltonian, Eq.(\ref{laplacian}),can be expressed now as:
\begin{equation}\label{}
H=\frac{p_3^2}{2m}-\frac{\hbar^2}{2m}(M^2-K)+\frac{-\hbar^2}{2m}(\frac{1}{h_1 h_2 h_3})(\partial_1\frac{h_2 h_3}{h_1}\partial_1+\partial_2\frac{h_1 h_3}{h_2}\partial_2)
\end{equation}
All the 3-dynamics is now contained in the first term, which is a Hermitian operator representing the contribution to the kinetic energy from the transverse 3-component of the
 momentum. As the particle is squeezed to the surface by the potential freezing the 3-degree of freedom, the contribution of the $\frac{p_3^2}{2m}$ to the energy can be safely taken as zero and so this term can be dropped from the Hamiltonian. Therefore, we have in the limit $q_3\rightarrow 0$ the surface Hamiltonian:
 \begin{equation}\label{}
H^{surf.}=\frac{-\hbar^2}{2m}\nabla'^2-\frac{\hbar^2}{2m}(M^2-K)
 \end{equation}
 where,
 \begin{equation}\label{laplacian prime}
 \frac{-\hbar^2}{2m}\nabla'^2=\frac{-\hbar^2}{2m}(\frac{1}{h_1 h_2 h_3})(\partial_1\frac{h_2 h_3}{h_1}\partial_1+\partial_2\frac{h_1 h_3}{h_2}\partial_2)\left.\right|_{q_3=a}
\end{equation}
This is the well-known \cite{Costa,Koppe} thin-layer quantization Hamiltonian for a spin zero particle on a curved surface. The  eigen functions of the above Hamiltonian $\psi(q_1,q_2)$ are normalized at the surface as :
\begin{equation}
  \int h_1 h_2 dq_1 dq_2 |\psi(q_1,q_2)|^2 =1
\end{equation}

\section{General Hermitian Hamiltonian of a spin zero particle coupled to an electromagnetic field on a surface in OCC's}
The Hamiltonian for a spin zero particle in a static electromagnetic field given by a vector potential $\vc{A}$ in an arbitrary gauge and  a scalar potential $V(\vc{q})$ in a 3D space spanned by an OCC system is :
\begin{eqnarray}\label{OCC H}
H&=&\frac{-\hbar^2}{2m}\nabla^2+\frac{i\hbar e}{m}(\frac{1}{h_1}A_1\partial_1+\frac{1}{h_2}A_2\partial_2+\frac{1}{h_3}A_3\partial_3)+\frac{i\hbar e}{2m}(\frac{1}{h_1 h_2 h_3})(\partial_1(h_2 h_3A_1)+\partial_2(h_3 h_1A_2)+\partial_3(h_1 h_2A_3))\\\nonumber
&+&\frac{e^2}{2m}\vc{A}\cdot\vc{A}+V(\vc{q})
\end{eqnarray}
$h_1 ,h_2$ and $h_3$ are the scale factors defined earlier.
The confinement of the particle to a surface embedded in 3D proceeds just as in the field-free case, again the confining potential is not shown in the Hamiltonian. The only difference is that in the presence of the electromagnetic field, we note that the physical gauge-covariant momentum is not the canonical one defined in Eq.(\ref{hermitian p}), rather it is the kinematical momentum operator :
\begin{equation}\label{kinetic momentum}
 \Pi_3=(p_3-eA_3)
\end{equation}
which is also the 3-velocity operator; recall $\dot{u}_3=\frac{[u_3,H]}{i\hbar}=\frac{\Pi_3}{m}$. Note that $p_3$ is now the Hermitian canonical momentum given by
 Eq.(\ref{hermitian p}).  We therefore write:
\begin{eqnarray*}\label{pi square}
 \frac{\Pi_3^2}{2m} &=& \frac{-\hbar^2}{2m}(\partial_3^2+\frac{\partial_3(h_1 h_2)}{h_1h_2}\partial_3)+\frac{\hbar^2}{2m}(\frac{\partial_3^2(h_1 h_2)}{2h_1 h_2}+\frac{(\partial_3(h_1 h_2))^2}{(2h_1 h_2)^2})\\\nonumber
  &+& \frac{i\hbar e}{m}A_3\partial_3++\frac{i\hbar e}{2m}(\partial_3A_3)+\frac{i\hbar e}{2m}(\frac{\partial_3(h_1 h_2)}{h_1h_2})A_3+\frac{e^2}{2m}A_3^2
\end{eqnarray*}

The above expression contains all the "3-terms" of the Hamiltonian, Eq.(\ref{OCC H}), including the 3-component of the gauge field and its 3-derivative. The second term in the above Hamiltonian is just the geometric kinetic energy term (see Eq.(\ref{geometric K}))
The fifth term is a coupling of the 3-component of the gauge field to the mean curvature of the surface \cite{shikakhwa and chair}:
\begin{equation}\label{geometric potential}
 -\frac{i\hbar g}{m}A_3(\frac{1}{2h_1 h_2 h_3}\partial_3(h_1 h_2))=\frac{i\hbar g}{m}A_3 M
\end{equation}
 So, we express the Hamiltonian now as :
 \begin{eqnarray}\label{OCC H2}
H&=&\frac{\Pi_3^2}{2m}-\frac{\hbar^2}{2m}(M^2-K)+\frac{-\hbar^2}{2m}(\frac{1}{h_1 h_2 h_3})(\partial_1\frac{h_2 h_3}{h_1}\partial_1+\partial_2\frac{h_1 h_3}{h_2}\partial_2)\\\nonumber
&+&\frac{i\hbar e}{2m}(\frac{1}{h_1 h_2 h_3})(\partial_1(h_2 h_3A_1)
+\partial_2(h_3 h_1A_2))
+\frac{i\hbar e}{m}(\frac{1}{h_1}A_1\partial_1+\frac{1}{h_2}A_2\partial_2)+\frac{e^2}{2m}(A_1^2+A_2^2)+V(\vc{q})
\end{eqnarray}
Note that all reference to the 3-degrees of freedom, including $A_3$ are now contained in $\Pi_3$ and do not appear elsewhere in the Hamiltonian. When the particle is squeezed to the surface by the potential thus freezing the transverse 3-degrees of freedom, then the gauge-invariant kinematical momentum  $\Pi_3$  can be dropped from the Hamiltonian, taking with it  all reference to the 3-dynamical quantities leaving us with the Hermitian gauge-covariant $A_3$-free surface Hamiltonian ($q_3\rightarrow a$):
 \begin{eqnarray}\label{Hermitian H}
H^{surf.}&=&\frac{-\hbar^2}{2m}\nabla'^2-\frac{\hbar^2}{2m}(M^2-K)
+\frac{i\hbar e}{2m}(\frac{1}{h_1 h_2 h_3})(\partial_1(h_2 h_3A_1)\\\nonumber
&+&\partial_2(h_3 h_1A_2))
+\frac{i\hbar e}{m}(\frac{1}{h_1}A_1\partial_1+\frac{1}{h_2}A_2\partial_2)+\frac{e^2}{2m}(A_1^2+A_2^2)+V
\end{eqnarray}
One can recast the above Hamiltonian in the explicitly gauge-covariant form :
\begin{align}\label{H covariant}
 H^{surf.}_{Herm.}&=\frac{-\hbar^2}{2m}((\hat{u}_1(\frac{1}{h_1}\partial_1-\frac{ie}{\hbar}A_1)+(\hat{u}_2(\frac{1}{h_2}\partial_2-\frac{ie}{\hbar}A_2))^2
 -\frac{\hbar^2}{2m}(M^2-K)+V\\\nonumber
   & =\frac{-\hbar^2}{2m}(\hat{u}_1D_1+\hat{u}_2D_2)^2-\frac{\hbar^2}{2m}(M^2-K)+V\\\nonumber
  &=\frac{-\hbar^2}{2m}(\vc{D'}\cdot\vc{D'})-\frac{\hbar^2}{2m}(M^2-K)+V
\end{align}
where we have defined the covariant derivatives $D_k=(\frac{1}{h_k}\partial_k-\frac{ie}{\hbar}A_k), k=1,2$, and the prime, again, denotes quantities on the surface and the absence of the 3-component. The above Hamiltonian is free from any reference to the transverse component; $A_3$, of the vector potential and the result is totally gauge-independent. This component is still there in our case, but is not "seen" at the surface ! This is in contrast to the results reported in previous works \cite{ferrari,Jensen1,Jensen2,Ortix1} where one has to carry out a gauge transformation taking it to zero  in order to eliminate it from the surface . Note also that the term given by Eq.(\ref{geometric potential})representing a problematic curvature contribution to the orbital magnetic moment which was reported to be  still present in the surface Hamiltonian when applying the standard thin layer quantization \cite{Jensen1,Jensen2} drops naturally in the current approach in agreement with the results in \cite{Ortix1}.   The key to our results is the physically-grounded argument that one has to set the gauge-covariant kinematical momentum transverse to the surface to zero as the particle is squeezed to the surface. What remains now is to carry out a gauge transformation $A_3\rightarrow A_3'=0$ - which is always possible-  to totally remove the now 'unseen' 3-component of the vector potential so as to express the magnetic field on the surface as $\vc{B}=\vc{\nabla}\times \vc{A'}|_{u_3=a}$ , with $ \vc{A'}=(A_1,A_2)$ now.
\begin{figure}[h!]
\includegraphics{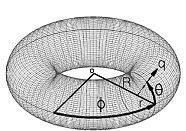}
\caption{Coordinates system $(\theta,\phi,q)$ near a torus surface. $R$ being the distance from the centre of the torus $O$ to the centre of the tube, $r$ is the radius of the tube and $q$ is the distance along the normal $\hat{n}$ }
\end{figure}

We now apply our general results developed above to a particle in three specific geometries; the surface of a cylinder, a sphere and a torus.
 Table \ref{table1} provides the h-factors, $p_3$'s and the geometric kinetic energies (GKE)for cylindrical, spherical and the torus (see figure) coordinates. The results for the GKE's coincide with the well-known ones reported in the literature \cite{exact,ferrari}.It is now straightforward to write down the Hermitian Hamiltonians in these three main OCC systems.
  \begin{table}[h]
  \centering
\begin{tabular}{|l|l|l|l|}
  \hline
  OCC&  h-factors  & $p_3$'s & GKE \\
   \hline
 Cylindrical& $q_1=\theta, q_2=z, q_3=r; h_\theta=r,h_z=h_r=1$ & $ p_r=-i\hbar(\partial_r+\frac{1}{2r})$ & $\frac{-\hbar^2}{8ma^2}$\\
   \hline
   Spherical& $q_1=\theta, q_2=\phi, q_3=r; h_\theta=r,h_\phi=r\sin\theta, h_r=1$ & $ p_r=-i\hbar(\partial_r+\frac{1}{r})$ & 0\\
     \hline
  Torus & $q_1=\theta, q_2=\phi, q_3=q; h_\theta=r+q,h_\phi=R+(r+q)\cos\theta, h_q=1$ & $ p_q=-i\hbar(\partial_q+\frac{(R+2(r+q)\cos\theta}{2(r+q)(R+r+q)\cos\theta})$ & $\frac{-\hbar^2 R^2}{8mr^2(R+r\cos\theta)^2}$\\
    \hline
\end{tabular}
  \caption{ The h-factors,$p_3$'s and GKE for cylindtical,spherical and torus coordinates. The expression for GKE is calculated by setting  $q_3=r=a$ at the surfaces of a cylinder and a sphere, and  $q_3=q=0$ at the surface of the torus}\label{table1}
\end{table}
 In cylindrical coordinates the Hermitian Hamiltonian on the  surface of a cylinder of radius $a$ is :
\begin{equation}\label{cyl}
H^{cyl.}=\frac{-\hbar^2}{2m}(\hat{\theta}D_\theta+\hat{z}D_z)^2+\frac{-\hbar^2}{8ma^2}
\end{equation}
with,
\begin{equation}\label{cov cyl}
 D_\theta=(\frac{1}{a}\partial_\theta-\frac{ie}{\hbar}A_\theta),D_z=(\partial_z-\frac{ie}{\hbar}A_z).
\end{equation}
Dropping the $z-$dependence, one gets the Hermitian Hamiltonian on a ring of radius $a$.On the surface of a sphere of radius $a$,the Hamiltonian takes the form:
\begin{equation}\label{sph}
H^{sph.}=\frac{-\hbar^2}{2m}(\hat{\theta}D_\theta+\hat{\phi}D_\phi)^2
\end{equation}
with,
\begin{equation}\label{cov sph}
 D_\theta=(\frac{1}{a}\partial_\theta-\frac{ie}{\hbar}A_\theta),D_\phi=(\frac{1}{a\sin\phi}\partial_\phi-\frac{ie}{\hbar}A_\phi).
\end{equation}
Finally, for a torus we have the Hermitian Hamiltonian on the surface of the torus ($q\rightarrow 0$) as:
\begin{equation}\label{torus}
  H^{tor}=\frac{-\hbar^2}{2m}(\hat{\theta}D_\theta+\hat{\phi}D_\phi)^2+\frac{-\hbar^2 R^2}{8mr^2(R+r\cos\theta)^2}
\end{equation}
with,
\begin{equation}\label{cov tor}
D_\theta=(\frac{1}{r}\partial_\theta-\frac{ie}{\hbar}A_\theta), D_\phi=(\frac{1}{R+r\cos\theta}\partial_\phi-\frac{ie}{\hbar}A_\phi)
\end{equation}
 Equations (\ref{cyl}-\ref{cov tor})are the most general expressions for the gauge-invariant Hermitian Hamiltonians of a spin zero particle in any magnetic field on the given surfaces.They are ready-to-use formulae that can be adapted to any vector potential. One needs only to express the given potential in the respective OCC and plug it into the relevant equation.

\section{Pauli equation for a spin one-half particle in an electromagnetic field with SOC on a curved surface}

The non-relativistic limit to $O(\frac{1}{c^2})$ of the Dirac equation in a static electromagnetic field is given as :
\begin{equation}\label{pauli equation}
H=\frac{1}{2m}(\vc{p}-e\vc{A})^2+V-\frac{e\hbar}{2m}\vc{\sigma}\cdot\vc{B}+\frac{e\hbar}{4m^2 c^2}\vc{\sigma}\cdot\vc{E}\times(\vc{p}-e\vc{A})
\end{equation}
Here, $\sigma$'s are the Pauli spin matrices, $\vc{A}$  the vector potential; $\vc{B}=\vc{\nabla}\times \vc{A}$  the magnetic field; and $\vc{E}$ is the electric field. The potential $V$ contains in addition to the scalar potential other $O(\frac{1}{c^2})$ terms, e.g.$\vc{\nabla}\cdot\vc{E}$...etc. The last term in the above Hamiltonian is the SOC. Note the appearance of the kinematical momentum operator $(\vc{p}-e\vc{A})$  in this term,a necessity for the $U(1)$ gauge-covariance of the SOC. It proves convenient to express the SOC using the $SU(2)$ non-Abelian gauge field formalism \cite{shikakhwa12,frohlich93,Jin.et.al-JPA,Dartora08}. We define:
\begin{equation}\label{W}
-gW_i^a=\frac{e\hbar}{2m}\epsilon_{iaj}E_j
\end{equation}
so that $\vc{W}=\vec{W}^a \sigma^{a}$ (or $\mathbf{W}_i=W_i^a \sigma^a$) with $i,a=1..3$. In this section we will use $a,b,c,...$ to refer to group indices and $i,j,k...$ to refer to space indices. The SOC term is now expressed as ( noting that $\vc{\nabla}\times\vc{E}=0$ and consequently $\vc{\nabla}\cdot\vc{W}=0$):
\begin{equation}\label{SOC}
\frac{e\hbar}{4m^2 c^2}\vc{\sigma}\cdot\vc{E}\times(\vc{p}-e\vc{A})=-\frac{g}{m}\vc{W}\cdot(\vc{p}-e\vc{A})
\end{equation}
Having in mind the relevant experimental  condensed matter systems, we take $\vc{E}$ and thus $\vc{W}$ as constant. In OCC, the above Hamiltonian then assumes the form:
\begin{eqnarray}\label{H Pauli OCC}
  H &=& \frac{-\hbar^2}{2m}\nabla^2+\frac{i\hbar e}{m}(\frac{1}{h_1}A_1\partial_1+\frac{1}{h_2}A_2\partial_2+\frac{1}{h_3}A_3\partial_3)+\frac{i\hbar e}{2m}(\frac{1}{h_1 h_2 h_3})(\partial_1(h_2 h_3A_1)+\partial_2(h_3 h_1A_2)+\partial_3(h_1 h_2A_3))\\\nonumber
   &+& \frac{e^2}{2m}\vc{A}\cdot\vc{A}+\frac{i\hbar g}{m}(W_1(\frac{1}{h_1}\partial_1-\frac{ie}{\hbar}A_1)+W_2(\frac{1}{h_2}\partial_2-\frac{ie}{\hbar}A_2)+W_3(\frac{1}{h_3}\partial_3-\frac{ie}{\hbar}A_3))
+V-\frac{e\hbar}{2m}\vc{\sigma}\cdot\vc{B}
\end{eqnarray}
Note that despite the introduction of the $SU(2)$ gauge field $\vc{W}$, the above Hamiltonian is \textit{not} $SU(2)$  gauge-covariant \cite{shikakhwa12}. We now rewrite this Hamiltonian separating  the transverse, i.e 3-degrees of freedom. Just as was done in the spin zero case, we use the definitions; Eqs.(\ref{hermitian p})and (\ref{kinetic momentum}), invoke  Eqs.(\ref{geometric K}) and (\ref{geometric potential}) with the latter read for  $W_3$ as well, to put the Hamiltonian in the form:
 \begin{eqnarray}\label{OCC H2}
H&=&\frac{\Pi_3^2}{2m}-\frac{\hbar^2}{2m}(M^2-K)+\frac{-\hbar^2}{2m}(\frac{1}{h_1 h_2 h_3})(\partial_1\frac{h_2 h_3}{h_1}\partial_1+\partial_2\frac{h_1 h_3}{h_2}\partial_2)\\\nonumber
&+&\frac{i\hbar e}{2m}(\frac{1}{h_1 h_2 h_3})(\partial_1(h_2 h_3A_1)
+\partial_2(h_3 h_1A_2))
+\frac{i\hbar e}{m}(\frac{1}{h_1}A_1\partial_1+\frac{1}{h_2}A_2\partial_2)+\frac{e^2}{2m}(A_1^2+A_2^2)\\\nonumber
&+&\frac{i\hbar}{m}W_3\Pi_3+\frac{i\hbar}{m}(W_1(\frac{1}{h_1}\partial_1-\frac{ie}{\hbar}A_1)+W_2(\frac{1}{h_2}\partial_2-\frac{ie}{\hbar}A_2))+\frac{i\hbar}{m}W_3M
+V-\frac{e\hbar}{2m}\vc{\sigma}\cdot\vc{B}
\end{eqnarray}
Note the appearance of the coupling between the 3-kinematical momentum and the 3-component of SO gauge field; the seventh term, as well as the coupling between the same component of the gauge field and the surface curvature; the ninth term. Now, in confining the particle to the surface, we drop $\Pi_3$ from the above Hamiltonian, and recall the result proven in \cite {shikakhwa and chair} that for a constant Cartesian field $\vc{W}$ , i.e. a constant SOC:
\begin{equation}\label{hermicity}
\frac{i\hbar g}{m}W_3M=-\frac{i\hbar g}{2m}(\frac{1}{h_1 h_2 h_3})(\partial_1(h_2 h_3W_1)
+\partial_2(h_3 h_1W_2))
\end{equation}
 thus reducing the Hamiltonian to the Hermitian gauge-covariant form:
 \begin{align}\label{H covariant}
 H^{Pauli}_{surf.}&=\frac{-\hbar^2}{2m}((\hat{u}_1(\frac{1}{h_1}\partial_1-\frac{ie}{\hbar}A_1)+(\hat{u}_2(\frac{1}{h_2}\partial_2-\frac{ie}{\hbar}A_2))^2
 -\frac{\hbar^2}{2m}(M^2-K)\\\nonumber
   &+\frac{i\hbar}{m}(W_1(\frac{1}{h_1}\partial_1-\frac{ie}{\hbar}A_1)+W_2(\frac{1}{h_2}\partial_2-\frac{ie}{\hbar}A_2))-\frac{i\hbar}{m}\vc{\nabla'}\cdot\vc{W'}
+V-\frac{e\hbar}{2m}\vc{\sigma}\cdot\vc{B}
\end{align}
Defining a generalized covariant derivative $ \mathfrak{D}_k=(\frac{1}{h_k}\partial_k-\frac{ie}{\hbar}A_k-\frac{ig}{\hbar}W_k), k=1,2 $, we can express this Hamiltonian in the explicitly $U(1)$ gauge-covariant form:
 \begin{align}\label{H covariant pauli}
 H^{Pauli}_{surf.}& =\frac{-\hbar^2}{2m}(\hat{u}_1\mathfrak{D}_1+\hat{u}_2\mathfrak{D}_2)^2-\frac{\hbar^2}{2m}(M^2-K)+V-\frac{e\hbar}{2m}\vc{\sigma}\cdot\vc{B}
 -\frac{g^2}{2m}\vc{W'}\cdot\vc{W'}\\\nonumber
  &=\frac{-\hbar^2}{2m}(\vc{\mathfrak{D}'}\cdot\vc{\mathfrak{D}'})-\frac{\hbar^2}{2m}(M^2-K)+V-\frac{e\hbar}{2m}\vc{\sigma}\cdot\vc{B}
  -\frac{g^2}{2m}\vc{W'}\cdot\vc{W'}
\end{align}
The last term is crucial as the Hamiltonian is not $SU(2)$ gauge-covariant as we have noted earlier. The above is the most general gauge-covariant Pauli Hamiltonian in a static electromagnetic field with constant SOC on a curved surface embedded in 3D space spanned by an OCC system.
\section{conclusions}

The approach presented in this work to write quantum mechanical Hamiltonians of a particle in a static electromagnetic field confined to a surface that is embedded in 3D space can be viewed as complementary to the well-established thin-layer quantization scheme. It infuses, in our opinion, strong physical intuition into this latter scheme. It is main argument is also simple: as the confining potential squeezes the particle to the surface, freezing the dynamics in the transverse direction, then the 'decoupling of the transverse dynamics' from the surface is to be achieved by setting the transverse component of the Hermitian and gauge-covariant physical momentum to zero in the Hamiltonian. This, not only reproduces the formal results of the thin-layer quantization scheme, it also provides two new results: The resulting surface Hamiltonian is totally free from the transverse component of the vector potential;$A_3$, and this is independent of any choice of the gauge, in contrast to earlier thin layer quantization-based approaches \cite{ferrari,Encinosa} where one needs to specify a given gauge that sets this  component to zero at the surface in order to decouple it from the surface dynamics or at least to remove it from the surface Hamiltonian.  It is interesting that in the present approach, this component of the vector potential just drops from the surface once the transverse kinematical momentum is dropped. It is not set to zero or to any specific value. As such, it is still 'out there' but is 'not seen' at the surface and can be -optionally- always gauged away to have $ \vc{A}=(A_1,A_2)$ in the whole space and not necessarily only at the surface. The second new result is the fact that the gauge-covariant, Hermitian and $A_3$-free surface Pauli Hamiltonian with constant SOC is constructed without any assumptions or approximations. \\
Eqs.(\ref{cyl}-\ref{cov tor}) provide the most general Hamiltonians for a particle in a static electromagnetic field on the surfaces of a cylinder,a sphere and a torus, respectively. As they stand, they are a plug-and-play formulae that can be enumerated by any vector potential.

Closing, we highlight one more interesting point. The extra term added to $-i\hbar\partial_3$ to render it Hermitian (see Eq.(\ref{hermitian p})) is just the mean curvature $M$ (see Eq.(\ref{geometric potential}))
\begin{equation*}
(\frac{1}{2h_1 h_2 h_3}\partial_3(h_1 h_2))=- M = g^{ij}\Gamma^3_{ij}
\end{equation*}
Here, $g^{ij}$ is the contravariant diagonal metric tensor of OCC, and  $\Gamma^3_{ij}=\frac{1}{2}g^{k3}(\partial_jg_{ik}+\partial_ig_{jk}-\partial_kg_{ij})$
is  \cite {Arfken} the Christoffel symbol of the second kind . Therefore,  the  Hermitian canonical momentum defined in Eq.(\ref{hermitian p}) can be expressed as:
\begin{equation*}
p_3=-i\hbar(\partial_3-M)
\end{equation*}
This general result suggests that one needs to add the curvature of the surface to the derivative of the transverse coordinate to construct the Hermitian or self-adjoint physical transverse momentum operator. It also points out the connection between the present approach and the thin-layer quantization scheme, suggesting at the same time that this approach can be applied for general -not necessarily orthogonal- curvilinear coordinates. To appreciate this last point, it is sufficient to look at Eq.(2) in reference \cite{Ortix1}. If one combines the last three terms of this equation, one gets our kinematical three momentum,$\Pi_3^2$, and the Schrodinger Hamiltonian in the equation reduces to our Hamiltonian, Eq.(\ref{OCC H2}).This, of course, signals that the present approach can be applied to general curvilinear coordinates.\\
After this work has been completed and put in the form of a preprint, the series of works \cite{Liu1,Liu2,Liu3,Liu4} were brought to our attention. In these works, the above expression for the momentum was noted and the term geometric momentum was coined. In the work \cite{Liu3}an expression analogous to Eq.(\ref{geometric K})was also derived for a general curvilinear coordinate. These results support our conjecture that our analysis can be indeed generalized to general curvilinear coordinates.

\end{document}